\documentclass[prb,onecolumn,superscriptaddress,showpacs]{revtex4}
\usepackage{psfrag,graphicx,amsmath}
\usepackage{hyperref}

\begin{document}

\title{Local density of states and Friedel oscillations around a non-magnetic impurity in unconventional density wave}

\author{Andr\'as V\'anyolos}
\email{vanyolos@kapica.phy.bme.hu}
\affiliation{Department of Physics, Budapest University of Technology and Economics, 1521 Budapest, Hungary}
\author{Bal\'azs D\'ora}
\affiliation{Department of Physics, Budapest University of Technology and Economics, 1521 Budapest, Hungary}
\affiliation{Max-Planck-Institut f\"ur Physik Komplexer Systeme, N\"othnitzer Strasse 38, 01187 Dresden, Germany}
\author{Attila Virosztek}
\affiliation{Department of Physics, Budapest University of Technology and Economics, 1521 Budapest, Hungary}
\affiliation{Research Institute for Solid State Physics and Optics, PO Box 49, 1525 Budapest, Hungary.}
\date{\today}

\begin{abstract}
We present a mean-field theoretical study on the effect of a single non-magnetic impurity in quasi-one dimensional unconventional
density wave. The local scattering potential is treated within the self-consistent $T$-matrix approximation. The local density of
states around the impurity shows the presence of resonant states in the vicinity of the Fermi level, much the same way as in
$d$-density waves or unconventional superconductors. The assumption for different forward and backscattering, characteristic to
quasi-one dimensional systems in general, leads to a resonance state that is double peaked in the pseudogap.  The Friedel
oscillations around the impurity are also explored in great detail, both within and beyond the density wave coherence length
$\xi_0$. Beyond $\xi_0$ we find power law behavior as opposed to the exponential decay of conventional density wave. The entropy
and specific heat contribution of the impurity are also calculated for arbitrary scattering strengths.
\end{abstract}

\maketitle

\section{Introduction}
The effects of a single magnetic and non-magnetic impurity on the properties of high temperature superconductors (HTSC) have
attracted considerable attention in the past few years both from the
theoretical,\cite{morr,jian,quiang,salkola,balatsky,balatsky-review} and more recently from the experimental
side.\cite{balatsky-review,pan,hudson,yazdani} This enthusiasm from the theory is mainly due to the fact, that electronic
scattering from a localized impurity produces qualitatively different electronic structure around the impurity in the $d$-wave
superconductor phase\cite{salkola,balatsky} (dSC) and in the alternative models proposed for the pseudogap phase of high-$T_c$
cuprates. These models include for example the $d$-wave density wave\cite{nayak-solo,laughlin,morr,quiang} (dDW), and the
superconducting phase-fluctuation scenario\cite{quiang,emery} (PF). As the local density of states (LDOS) is proportional to the
differential tunneling conductance, scanning tunneling microscopy (STM) measurements are in principle able to distinguish between
the different available scenarios due to their spatial and energy resolution capability. Indeed, recent STM spectroscopy performed
on Bi$_2$Sr$_2$CaCu$_2$O$_{8+\delta}$ HTSC below the superconducting transition temperature $T_c$ confirmed the existence of the
impurity induced quasiparticle subgap resonance close to the Fermi energy, anticipated in a dSC.\cite{pan,hudson,yazdani}

The understanding of the microscopic origin of the pseudogap (PG) phase of HTSC represents a formidable challenge for both theory
and experiment. Many experiments have provided evidence for unusual normal state behavior in underdoped cuprates. One of the most
striking is definitely the observed gaplike feature of the quasiparticle density of states (DOS) above $T_c$, with depleted
electronic states around the Fermi energy.\cite{renner} It was shown lately, that independently of the origin of the PG state,
such a DOS is alone sufficient to produce a resonant impurity state close to the Fermi level in the presence of a local
non-magnetic impurity.\cite{kruis} The development of this single quasiparticle peak in the pseudogap violates particle-hole
symmetry. Similar conclusions were reached in the dDW\cite{morr,quiang} and PF\cite{quiang} models too, although in the latter case
the subgap resonance was found to be double peaked.

The study of electronic scattering from a localized perturbation however is not limited to HTSC physics. In fact, this issue was
on agenda already in the late sixties, in the context of conventional BCS superconductivity.\cite{shiba,machida,hotta} It was
found that in $s$-wave superconductor both magnetic,\cite{shiba} and resonant non-magnetic\cite{machida} impurities produce true
bound states inside the energy gap. In addition to that, conventional charge density waves (CDW) were also investigated from this
aspect,\cite{tutto,cheng,zawatutto,hansen} and intragap bound states were obtained here as well.

In this theoretical study, our aim is to explore the effects of a single non-magnetic impurity in quasi-one dimensional
unconventional density wave (UDW). As the scattering is spin independent, our theory and conclusions are applicable to both
unconventional charge density wave (UCDW) and unconventional spin density wave (USDW). An UDW is a density wave with momentum
dependent gap $\Delta(\mathbf{k})$,\cite{balazs-modern,balazs-sdw} the average value of which over the Fermi surface
vanishes. Thus these systems lack spatial modulation of either charge or spin density that could be observed in principle by
conventional means (x-ray or NMR). Though this sort of hidden order makes experimental detection more complicated, still a great
deal of effort has been done for example in organic conductor
$\alpha$-(BEDT-TTF)$_2$KHg(SCN)$_4$.\cite{andres,mori,kartsovnik,basletic,fujita,pouget} The low temperature phase of this salt
has been explained rather well by a quasi-one dimensional UCDW, see Ref.~[\onlinecite{balazs-modern}] and references
therein. Potential scattering from impurities with finite concentration has been investigated in UDW in the weak scattering Born
limit,\cite{balazs-born} in the unitary limit,\cite{balazs-unitary} and more recently in an extended scheme valid for arbitrary
scattering amplitudes\cite{nca-impurity}. To the best of our knowledge however, the single impurity problem in UDW has not been
addressed so far. Therefore we study this issue here. We explore the electronic structure and Friedel oscillations around the
impurity. Besides, the specific heat contribution is calculated as well for arbitrary scattering strength.

The paper is organized as follows: in Section 2 we present the formalism of the $T$-matrix approach. In Sections 3-5 the formalism
is applied to quasi-one dimensional unconventional density wave in order to obtain numerical and analytical results for the local
density of states around the impurity, the induced Friedel oscillations, and for the thermodynamics, respectively. Finally,
Section 6 is devoted to our summary and conclusion. In Appendix A further useful results are collected, those we make use of
throughout the article.

\section{Formalism}
We begin with the mean-field Hamiltonian describing pure quasi-one dimensional density waves\cite{gruner-book,balazs-sdw}
\begin{equation}
H_0=\sideset{}{'}\sum_{\mathbf{k},\sigma}\xi(\mathbf{k})(a_{\mathbf{k},\sigma}^\dag a_{\mathbf{k},\sigma}-a_{\mathbf{k-Q},\sigma}^\dag
a_{\mathbf{k-Q},\sigma})+\Delta_\sigma(\mathbf{k})a_{\mathbf{k},\sigma}^\dag 
a_{\mathbf{k-Q},\sigma}+\Delta_\sigma^*(\mathbf{k})a_{\mathbf{k-Q},\sigma}^\dag 
a_{\mathbf{k},\sigma},\label{meanfieldhamilton}
\end{equation}
where $a_{\mathbf{k},\sigma}^\dag$ and $a_{\mathbf{k},\sigma}$ are, respectively, the creation and annihilation operators for an
electron in a single band with momentum $\mathbf{k}$ and spin $\sigma$. The prime on the summation means that it is restricted to
the reduced Brillouin zone as $|k_x-k_F|<k_c$, where $k_c$ is the momentum cutoff. The best nesting vector is given by
$\mathbf{Q}=(2k_F,\pi/b,\pi/c)$, with $k_F$ being the Fermi wave number. Our system is based on an orthorhombic lattice with
lattice constants $a,b,c$ towards directions $x,y,z$. The system is highly anisotropic, the kinetic energy spectrum of the
particles linearized around the Fermi surface reads as
\begin{equation}
\xi(\mathbf{k})=v_F(k_x-k_F)-2t_b\cos(bk_y)-2t_c\cos(ck_z),\label{kinetic}
\end{equation}
and the quasi-one dimensional direction is the $x$ axis. For (U)CDW, the density wave order parameter $\Delta_\sigma(\mathbf{k})$
is even, while in (U)SDW it is an odd function of spin index. In the case of conventional density waves the order parameter is
constant on the Fermi surface.\cite{gruner-book} As opposed to this, for unconventional condensates it depends on the
perpendicular momentum and has different values at different points on the Fermi surface. The precise $\mathbf{k}$-dependence is
determined by the matrix element of the electron-electron interaction through the gap equation.\cite{balazs-sdw}

In the presence of a local impurity, the order parameter is no longer homogeneous in real space. A self-consistent calculation
with the interaction included results in a spatially varying $\Delta(\mathbf{r})$,\cite{shiba} where the deviation from the bulk
value, arising from pair-breaking effect, is confined to the vicinity of the impurity itself. The Zn substitution in cuprates is a
good example of this. However, the phenomena discussed here are not expected to be altered significantly by introducing position
dependent corrections.\cite{shiba} Thus, we completely ignore this effect and stick with the homogeneous solution, the momentum
dependence of which is chosen as $\Delta(\mathbf{k})=\Delta e^{i\phi}\sin(bk_y)$,\cite{balazs-sdw} where the phase $\phi$ is
unrestricted due to incommensurability.

The interaction of electrons with the single non-magnetic impurity placed at the origin is described by the Hamiltonian
\begin{equation}
H_1=\frac1V\sideset{}{'}\sum_{\mathbf{k,q},\sigma}
\begin{pmatrix}
 a_{\mathbf{k+q},\sigma}\\
 a_{\mathbf{k+q-Q},\sigma}
\end{pmatrix}^\dag
\begin{pmatrix}
U(0) & U(\mathbf{Q})\\
U(\mathbf{Q}) & U(0)
\end{pmatrix}
\begin{pmatrix}
 a_{\mathbf{k},\sigma}\\
 a_{\mathbf{k-Q},\sigma}
\end{pmatrix}.\label{impurity}
\end{equation}
Here the summation over momentum transfer $\mathbf{q}$ is restricted to small values, $V$ is the sample volume and we have
neglected the small $\mathbf{q}$-dependence of the matrix elements (i.e. the Fourier components of the impurity potential). This
is because we are dealing with quasi-one dimensional density wave systems, whose Fermi surface consists of two almost parallel
sheets at the points $\pm k_F$. Consequently, two relevant scattering amplitudes can be distinguished: the forward $U(0)$ and the
backward scattering parameter $U(\mathbf{Q})$, respectively. We believe that we can capture the essence of physics with this
approximation. This is indeed the case, at least as long as only static quantities, single particle properties and thermodynamics
is concerned. Investigating the effect of pinning, and dynamic properties such as the sliding density wave however, requires the
inclusion of the small $\mathbf{q}$ contribution as well.\cite{balazs-threshold,balazs-magneticthreshold}

It is worthwhile to emphasize at this point, that the allowance for different forward and backscattering in Eq.~\eqref{impurity}
constitutes a more realistic impurity physics than the usage of a somewhat artificial point-like scalar impurity
($U(\mathbf{r})=U\delta(\mathbf{r}))$, with $U(0)=U(\mathbf{Q})\equiv U$. Nevertheless, this restricted impurity model is widely
used in literature concerning the single impurity problem in conventional charge density waves,\cite{tutto,cheng} unconventional
superconductors with $d$-wave pairing symmetry,\cite{salkola,balatsky} and more recently in the context of the pseudogap phase of
HTSC.\cite{morr,kruis,jian,quiang} We will see shortly however, that the generalized approach with $U(0)\ne U(\mathbf{Q})$ will
reveal the true double peaked nature of the subgap resonance in UDW. These two virtually bound states with finite lifetime
correspond to the infinitely sharp, and therefore well-defined bound states found in fully gapped CDW.\cite{zawatutto} We suspect,
that similarly to quasi-one dimensional density waves, analogous results for the counterpart resonance peak could be obtained in
quasi-two dimensional $d$-density wave.

The main part of the paper is devoted to calculating the effect of impurity on the spectral function $A(\mathbf{r},\omega)$ (in
other words the local density of states), and the total electron density $n(\mathbf{r})$. These quantities are directly related to
the single particle Green's function. In view of the change in thermodynamic potential $\delta\Omega$, caused by the interaction,
we will also determine later on the specific heat contribution of dilute impurities at arbitrary scattering strengths. As the
interaction Hamiltonian in Eq.~\eqref{impurity} is quadratic in fermion operators, $\delta\Omega$ is thus related to the Green's
function as well. According to all these, our first goal is to calculate the single particle propagator dressed by the scatterer
to infinite order. In view of this, the rest can be obtained in a relatively straightforward manner. Hereafter we drop spin
indices since they are irrelevant for our discussion on the effect of potential scattering, and for the sake of simplicity
consider spinless fermions but our conclusions apply to both unconventional spin and charge density wave.

In the usual way the electron field operator can be split into left ($L$) and right ($R$) moving parts as
$\Psi(\mathbf{r})=\Psi_L(\mathbf{r})+\Psi_R(\mathbf{r})$,\cite{tutto} where
\begin{equation}
\Psi_\alpha(\mathbf{r})=\frac{1}{\sqrt{V}}\sideset{}{'}\sum_{\mathbf{k}}e^{i\mathbf{kr}}
a_\mathbf{k},\qquad\alpha=R(=+1),\,L(=-1),\label{fieldoperator}
\end{equation}
and the prime on the summation means the constraint $|k_x-\alpha k_F|<k_c$ in accordance with
Eq.~\eqref{meanfieldhamilton}. Choosing a sharp cutoff is somewhat artificial and obviously model dependent. Nevertheless, we
shall follow the treatment developed for conventional density wave in Ref.~[\onlinecite{tutto}], and the cutoff will be taken to
infinity whenever it does not affect the essence of physics. In this respect, most quantities we evaluate in the paper are
insensitive to the applied cutoff.

The definition of Green's function including the effect of impurity is
\begin{equation}
G_{\alpha\beta}(\mathbf{r,r'};\tau)=-\langle T_\tau[\Psi_\alpha(\mathbf{r},\tau)\Psi_\beta^\dag(\mathbf{r'})]\rangle_H,
\label{definition}
\end{equation}
where $H$ refers to the total Hamiltonian $H_0+H_1$. Applying standard equation-of-motion technique, Dyson's equation can be given
in a matrix form for the Matsubara components as\cite{tutto}
\begin{equation}
G_{\alpha\beta}(\mathbf{r,r'};i\omega_n)=G^0_{\alpha\beta}(\mathbf{r,r'};i\omega_n)+
G^0_{\alpha\gamma}(\mathbf{r},0;i\omega_n)t_{\gamma\delta}G_{\delta\beta}(0,\mathbf{r'};i\omega_n),\label{dyson}
\end{equation}
where $t_{\gamma\delta}=U(0)\delta_{\gamma\delta}+U(\mathbf{Q})\delta_{\gamma,-\delta}$, the zero superscript stands for the bare
propagator, and summation is applied over indices occurring twice. Now the spectral function, the particle density and the change
in the grand canonical potential are explicitly given as follows
\begin{gather}
A(\mathbf{r},\omega)=-\frac{1}{\pi}\text{Im}\sum_{\alpha\beta}G_{\alpha\beta}(\mathbf{r,r};i\omega_n\to\omega+i0),\label{localdos}\\
n(\mathbf{r})=\sum_{\alpha\beta}G_{\alpha\beta}(\mathbf{r,r};\tau=-0),\label{density}
\end{gather}
and finally, $\delta\Omega$ is calculated from the coupling constant integral as
\begin{align}
\delta\Omega&=\int_0^1 d\lambda\,\lim_{\tau\to-0}\left\{U(0)[G_{RR}(0,0;\tau)_\lambda+G_{LL}(0,0;\tau)_\lambda]+
U(\mathbf{Q})[G_{RL}(0,0;\tau)_\lambda+G_{LR}(0,0;\tau)_\lambda]\right\}.\label{domega}
\end{align}

In the following sections we give detailed description of the quantities above in UDW, and will compare them to those of a fully
gapped conventional DW.\cite{tutto,cheng,zawatutto,hansen} Besides, emphasis will be put also on the fact, that the local density
of states around the scalar impurity in UDW, in particular its intragap structure, greatly resembles that of the dDW phase
obtained in Refs.~[\onlinecite{morr,jian,quiang}].

\section{Local density of states}
In this section we present the results for the spectral function making use of Eq.~\eqref{localdos}. Dyson's equation for the
Green's function can be readily solved as
\begin{equation}
G_{\alpha\beta}(\mathbf{r,r'};i\omega_n)=G^0_{\alpha\beta}(\mathbf{r,r'};i\omega_n)+G^0_{\alpha\gamma}(\mathbf{r},0;i\omega_n)
T_{\gamma\delta}(i\omega_n)G^0_{\delta\beta}(0,\mathbf{r'};i\omega_n),\label{dysonsolution}
\end{equation}
where
\begin{equation}
T_{\gamma\delta}(i\omega_n)=\left\{t[1-G^0(0,0;i\omega_n)t]^{-1}\right\}_{\gamma\delta}\label{tmatrix}
\end{equation}
is the $T$-matrix. Equation~\eqref{dysonsolution} provides us the general dependence of $G_{\alpha\beta}(\mathbf{r,r'};i\omega_n)$
on $\mathbf{r}$ and $\mathbf{r'}$, in the followings however, according to Eqs.~(\ref{localdos}-\ref{domega}) we consider only
diagonal components (in real space). Furthermore, in a quasi-one dimensional UDW besides the chain direction $x$, in which the
model is continuous, perpendicular spatial dimensions are present as well offering the possibility for momentum dependent
gap.\cite{balazs-sdw,balazs-modern} This is to be contrasted with the strictly one-dimensional nature of normal
CDW.\cite{balazs-friedel} As the specific $\mathbf{k}$-dependence of the density wave order parameter was chosen to be
$\Delta(\mathbf{k})\sim\sin(bk_y)$, the relevant perpendicular direction is $y$, and the model remains discrete in this
variable. Consequently we take $\mathbf{r}=(x,mb,0)$, where $b$ is the corresponding lattice constant and $m$ is an integer
indexing parallel chains.

\begin{figure}[t]
\caption{\label{fig:ldos} The dimensionless local density of states $A(\mathbf{r},\omega)/N_0$ is shown in UDW versus energy and
  position measured from the scatterer for $m=0$ (top left), $m=\pm1$ (top right), $m=\pm2$ (bottom left) and $m=\pm3$ (bottom
  right). For the numerics $N_0U(0)=0.2$, $N_0U(\mathbf{Q})=0.7$, $v_Fk_F/\Delta=v_Fk_F/t_b=10$ and $\phi=\pi/3$ were applied. The
  black lines on the top left figure were calculated from $k_Fx=n\pi v_Fk_F/(\omega+v_Fk_F)$ with $n=5$, 6, 7 and 8 (from bottom
  to top).}
\end{figure}

It is clear from the structure of the $T$-matrix, that while analytically continuing to real frequencies as prescribed in
Eq.~\eqref{localdos}, poles can emerge at certain well defined energies that describe bound states localized at the impurity
site. This phenomenon is well understood in a normal metal for quite a long time.\cite{clogston} There, the binding energy of
these states are either above the upper or below the lower edge of the band. Bound states can also show up in conventional density
wave, though the situation in this case is more complex because additional states can develop in the gap as
well.\cite{tutto,cheng,zawatutto,hansen} We will see in a little while that in unconventional density wave, though being
gapless,\cite{balazs-modern} localized intragap states develop near the Fermi energy (see the region $|\omega|<\Delta$ in
Figs.~\ref{fig:ldos} and \ref{fig:ldossmall}). Albeit these states closely resemble to the corresponding results in fully gapped
charge or spin density wave, their widths in energy are not infinitely sharp and therefore the identification as true bound states
is not adequate. Rather we call them virtual or resonance states. In normal metal, this sort of accumulation of states in a
broader energy range is referred to as a virtual state as well.\cite{clogston} On the other hand, the notion of a resonance state
is widely used in the terminology of unconventional ($d$-wave, etc.) superconductivity,\cite{balatsky,salkola} and more recently
in the context of the pseudogap phase of high-$T_c$ cuprates.\cite{kruis,morr,jian}

Making use of the solution of Dyson's equation and the results obtained in Appendix A regarding the bare Green's function, we
numerically determined and plotted $A(\mathbf{r},\omega)$ for UDW in Fig.~\ref{fig:ldos}. The four panels show the energy and
position dependence of the spectral function around the impurity. For example the top left panel corresponds to the $m=0$ chain,
that is the one where the impurity resides. The other three demonstrate the effect of the scatterer on the neighboring chains with
$m=\pm1$, $m=\pm2$ and $m=\pm3$, respectively. It is immediately clear from the figures that the presence of the local
perturbation violates particle-hole symmetry. This is a common feature of potential scattering and is familiar in
dSC\cite{salkola,balatsky} and dDW\cite{morr,jian} as well. In addition to that, the local electronic structure exhibits quite
similar patterns on every chain. The fine details and features of these patterns can be nicely separated into three distinct
components, each having its own microscopic origin. These will be studied in the followings:

(i) Perhaps the most apparent flavor of the images in Fig.~\ref{fig:ldos} are those curved stripes or waves that are becoming ever
denser at high energies. They are essentially nothing else but the electronic wavefunctions and can be obtained even in a strictly
one-dimensional metal. A simple de Broglie picture can already account for the observed periodicity along the chain:
$\lambda=2\pi/p$, where $p=(\omega+v_Fk_F)/v_F$ is the momentum of the electron with energy $\omega$ measured from the chemical
potential in a linearized band. The very same electronic waves were found in the spectral function of one-dimensional conventional
CDW.\cite{balazs-friedel} Namely, in Ref.~[\onlinecite{balazs-friedel}] the effect of open boundary on CDW was studied. Among
others it was found that the position of zeros in the STM image is determined by $k_Fx=n\pi v_Fk_F/(\omega+v_Fk_F)$, with $n$ a
natural number. This result is in complete agreement with our simple reasoning based on de Broglie formula. Each stripe can
therefore be assigned a natural number $n$ (see especially the top left panel with the added curves), though they do not
necessarily indicate zeros in UDW. This is because the pattern in the present case depends (weakly) on the scattering amplitudes
too, and an exact agreement is achieved only in the limit $U(0)=U(\mathbf{Q})\to\infty$ corresponding to the case of open
boundary.

\begin{figure}[t]
\caption{\label{fig:ldossmall}Left panel: the local density of states in UDW is plotted on the $m=0$ chain, enlarged around the
  Fermi energy to better show the resonant states in the pseudogap. The other parameters are the same as on
  Fig.~\ref{fig:ldos}. Right panel: the same quantity for fixed distances from the scatterer, $k_Fx=0.9$ (dashed), 1.8
  (dashed-dotted) and 4 (solid).}
\end{figure}

(ii) An other interesting feature of the STM images is that there is a modulated behavior along the chains with a much larger
wavelength than $\lambda$. Namely, $A(\mathbf{r},\omega)$ ``periodically'' takes on the metallic value $N_0$, in other words
exhibits a beat. This property arises from the position dependence of the zeroth order Green's function, see Appendix A. One can
check either from the numerics or from the analytical calculation, that on chain $m$ whenever $J^2_m(2x/\xi)=0$, the LDOS is that
of a normal metal. Here $J_m(z)$ is the Bessel function of the first kind and $\xi=v_F/t_b$ is the characteristic lengthscale
originating from finite interchain coupling. In quasi-one dimensional density wave materials $t_b$ is usually of the order of the
energy gap $\Delta$, and this leads to a $\xi$ that is comparable to the DW coherence length $\xi_0=v_F/\Delta$, in any case it is
much larger than atomic distances. This modulated behavior related to the zeros of the Bessel function is clearly missing in
strictly one-dimensional models and is a speciality of real quasi-one dimensional systems, let it be either a normal metal, a CDW
or even UDW.

(iii) The most important property of the electronic structure, at least from the UDW point of view, is definitely the low energy
behavior around the Fermi level. The subgap structure of LDOS, as shown enlarged in the left panel of Fig.~\ref{fig:ldossmall}, is
qualitatively different from that of a fully gapped DW.\cite{balazs-friedel,tutto,cheng,zawatutto,hansen} Namely, in UDW
considerable amount of spectral weight is accumulated in the intragap regime in the form of two virtually bound quasiparticle
states. The energies of these impurity states are determined by the poles of the $T$-matrix
\begin{equation}
[1-U_\pm g_1(\Omega_\pm)]^2+[U_\pm g_2(\Omega_\pm)]^2=0,\label{center}
\end{equation}
where $U_\pm=U(0)\pm U(\mathbf{Q})$. In principle, the solutions of Eq.~\eqref{center} are complex, indicating the resonant nature
of the virtual states. Making use of the explicit forms of $g_{1,2}$ given by Eqs.~\eqref{g1new} and \eqref{g2new}, the energies
$\Omega_\pm'$ and the decay rates $\Omega_\pm''$ are obtained as
\begin{equation}
\Omega_\pm\equiv\Omega_\pm'-i\Omega_\pm''=-\frac{\Delta}{N_0U_\pm}\frac{1}{\ln(4N_0|U_\pm|)}
\left(1+i\frac\pi2\frac{\text{sign}(U_\pm)}{\ln(4N_0|U_\pm|)}\right),\label{virtual}
\end{equation}
where we have assumed the impurity scattering to be close enough to the unitary limit so that the result can be computed to
logarithmic accuracy with $\ln(4N_0|U_\pm|)\gg1$. It is only in this limit that the two bound states are well defined with
$\Omega_\pm''\ll|\Omega_\pm'|$. It is also easy to see, that the finite lifetime results from the finite density of states in the
subgap coming from nodal quasiparticles ($A^0(\omega)=-(2/\pi)g_2(\omega)\sim|\omega|$ for small $\omega$). In contrast to
this, in conventional DW with constant gap the binding energies of the corresponding impurity states are purely real leading to
infinitely sharp resonances and well-defined undamped states. In particular, for example if forward scattering is ignored
($U(0)=0$), two symmetrically placed states with opposite energies are found.\cite{tutto,hansen} Equation~\eqref{virtual} and the
presence of impurity induced quasiparticle states in the UDW gap are in precise agreement with that found in $d$-wave
superconductor,\cite{salkola,balatsky} and in $d$-density wave.\cite{kruis,morr} However, in dDW only one such impurity state has
been found, which is due to the limitation of the strictly point-like impurity potential applied there. Indeed, for such a
potential $U_-=0$ and the pole structure of the $T$-matrix exhibits a single resonance only. Note also, that in dSC and in the PF
scenario\cite{quiang} of the PG phase, a double peaked resonance is found: one state on both positive (electron) and negative
(hole) biases. This structure arises from the particle-hole mixing, an essential feature of pairing in SC, and has nothing to do
with different forward and backscattering.

In quasi-one dimensional (U)DW systems the fine tuning of chemical potential (doping) does not play such an important role as for
instance in quasi-two dimensional dDW. It is because it leaves the nesting property unaffected, as $\mu$ can be incorporated in
$k_F$. Consequently, the energies of the intragap resonance peaks are not affected by doping, as opposed to that found in dDW,
where it scales with $\mu$.\cite{quiang} In this respect UDW behaves much the same way as a dSC\cite{quiang}: there due to the
pairing mechanism the quasiparticles are always excited with respect to the Fermi energy, leading to impurity induced states with
energy pinned almost on the Fermi surface and essentially unaffected by $\mu$.

\section{Friedel oscillation}
This section is devoted to the analysis of density oscillations caused by the impurity in the Born limit. The investigation of
this issue is motivated by the findings of Ref.~[\onlinecite{tutto}], valid in one-dimensional conventional charge density
wave. Namely, it was found, that at zero temperature below the coherence length $\xi_0=v_F/\Delta$, the charge density around the
impurity is just the sum of the contributions corresponding to the CDW and the Friedel oscillations. Beyond $\xi_0$ however, as
the necessary electron-hole pairs with energy smaller than $2\Delta$ are not available, Friedel oscillations essentially cannot
build up. The exponential decay of the oscillations at this lengthscale was interpreted as a tunneling effect.

Our aim in this section is to perform an analogue but finite temperature calculation in UDW, and find the total density below and
beyond the coherence length. This enterprise constitutes a more difficult problem than the determination of the one-dimensional
$T=0$ CDW response performed in Ref.~[\onlinecite{tutto}] because of the following three reasons: (i) As UDW is a quasi-one
dimensional structure, the finite interchain coupling has to be taken into account as well. This introduces an additional
lengthscale $\xi=v_F/t_b$ in the analytical calculation besides $\xi_0$ and the atomic distances $v_F/D$, $b$.  Consequently the
Friedel oscillations will become essentially two-dimensional and the chains with index $m\ne0$ will be affected by the impurity
too. (ii) In contrast to fully gapped DW, an UDW is characterized by a momentum dependent order parameter and this extra
$\mathbf{k}$-dependence makes explicit calculation more cumbersome. (iii) At finite temperature $T$ the temperature itself
introduces a characteristic length: $\xi_1=v_F/T$. Although the treatment of finite $T$ obviously leads to extra complications in
the actual calculation, we try to incorporate its effect in our theory. We will see shortly that it is indeed important because it
leads to qualitative changes compared to the zero temperature results.

Since in UDW no periodic modulation of either charge or spin density is present, we expect robust Friedel oscillations showing up
below $\xi_0$. On the other hand, as UDW is gapless, and nodal excitations are available with arbitrary small energy, we expect
the oscillations beyond the coherence length to exhibit power law behavior as opposed to exponential decay.

\subsection{Normal metal}
To get started, we first present the results for the density oscillations obtained for a quasi-one dimensional normal metal in the
Born limit, where $T_{\gamma\delta}(i\omega_n)=t_{\gamma\delta}$. Using Eq.~\eqref{density} and the Matsubara versions of the bare
Green's function given in Appendix A, for $|x|\gg v_F/D$ one obtains
\begin{equation}
n(x,m)=n_0-n_0N_0U(\mathbf{Q})\pi(-1)^mJ_m^2\left(\frac{2x}{\xi}\right)P\left(\frac{2\pi|x|}{\xi_1}\right)
\frac{\cos(2k_Fx)}{2k_F|x|},\label{bornmetal}
\end{equation}
where $n_0=k_F/(\pi bc)$ is the homogeneous density, $P(z)=z\sinh^{-1}(z)$ and $J_m(z)$ is the Bessel function of the first
kind. This is a very instructive result and it is worth stopping here for a moment and analyze it in detail. The first thing it
tells us is that in a normal metal in the Born limit only backscattering contributes to Friedel oscillations as the forward
scattering amplitude drops out from the calculation. It is also clear from Eq.~\eqref{bornmetal} that the different lengthscales
are factorized in the sense that they appear in different terms of a product. The last term is the familiar Friedel oscillation
with periodicity $2k_F$ and algebraic asymptotics ($\sim|x|^{-1}$). This behavior is known to be the consequence of the sharp
Fermi surface of normal metal at zero temperature. In other words, from a more physical point of view, the long range oscillations
develop because it is not possible to construct a smooth function out of the restricted set of wave vectors $|k|<k_F$. Here, in
our finite temperature result we find that this oscillating term is modulated by a smooth envelope $P(z)$, that in the zero
temperature limit correctly simplifies to $P(z\to0)=1$. On the other hand, at arbitrary small but finite temperature the
long-range behavior is replaced by an exponential decay according to $P(z\gg1)\approx2ze^{-z}$.\cite{fetter} This qualitative
change arises from the fact that at finite temperature the Fermi surface is smeared over a thickness $T$ in energy and the
electronic distribution function becomes smooth and analytic.

As to the effect of impurity on the parallel chains, it has been taken into account in Eq.~\eqref{bornmetal} by the Bessel
function. In the extreme limit of decoupled one-dimensional chains, where $t_b\to0$, one readily finds
$J_m(2x/\xi)\to\delta_{m0}$, indicating the fact that screening takes place on the chain only where the impurity resides and all
the others are completely unaffected. At finite interchain coupling the square of the Bessel function serves as a modulating
function: due to its quasiperiodic zeros it results in a beat in the induced charge density with wavelength
$\lambda\approx\pi\xi$, that is certainly much larger than that of the Friedel oscillation, $\pi/k_F$.

\subsection{Conventional CDW}
The corresponding formula for the impurity induced charge response in a quasi-one dimensional conventional CDW can also be
obtained in closed form at distances larger than the atomic length scale. Below the transition temperature the order parameter is
finite, and one can easily show that in this case the Born calculation acquires an extra contribution from forward scattering
processes. However, it is small and does not exhibit the relevant $2k_F$ periodicity characteristic to Friedel oscillation. We
therefore omit it here and concentrate on the effect of backscattering only. With all this, the analogue of Eq.~\eqref{bornmetal}
in CDW at zero temperature reads as
\begin{equation}
n(x,m)=n_0-(-1)^mn_1\cos(2k_Fx+\phi)-
n_0N_0U(\mathbf{Q})\pi(-1)^mJ_m^2\left(\frac{2x}{\xi}\right) F\left(\frac{2|x|}{\xi_0}\right)
\frac{\cos(2k_Fx)}{2k_F\xi_0},\label{borncdw}
\end{equation}
where
\begin{equation}
F(z)=2K_1(z)-2e^{i\phi}\cos(\phi)\left(\frac{\pi}{2}-zK_0(z)-\frac{\pi}{2}z\left(L_1(z)K_0(z)+L_0(z)K_1(z)\right)\right).
\label{nagyf}
\end{equation}
Here $n_1=\Delta/|g|$ with $g$ being the density wave coupling constant. Furthermore $K_n(z)$ and $L_n(z)$ are, respectively, the
modified Bessel function of the second kind and the Struve function. The first two terms together in Eq.~\eqref{borncdw} is the
usual CDW charge modulation. In a pure CDW without impurity, the phase $\phi$ of the order parameter is unrestricted due to
incommensurability. However, introducing the perturbing potential the phase and therefore the overall position of the condensate
gets pinned to $\phi(U(\mathbf{Q})>0)=0$, or $\phi(U(\mathbf{Q})<0)=\pi$.\cite{tutto} In either case $F(z)$ in Eq.~\eqref{nagyf}
remains the same, and within the CDW coherence length it simplifies to $F(z\ll1)\approx2z^{-1}$. In this region therefore the
contribution of the impurity to the Friedel oscillations is precisely that of the normal metal. On the other hand at distances
much larger than $\xi_0$ we have $F(z\gg1)\approx\sqrt{2\pi}z^{-3/2}e^{-z}$, and exponential decay is obtained in agreement with
Ref.~[\onlinecite{tutto}], see also Fig.~\ref{fig:friedel}.

Strictly speaking, Eq.~\eqref{borncdw} is valid only at absolute zero temperature. However, its validity still holds for low
temperatures where $\beta\Delta(T)=\xi_1/\xi_0\gg1$. The justification of this statement is as follows: As we have just seen, at
large distances exponential decay is obtained with characteristic length $\xi_0$. This behavior is very similar to the effect of
finite $T$ in normal metal. The obvious parallelism can be easily understood by remembering the fact that in a conventional CDW
with constant gap the Fermi surface is smeared over an energy width $\Delta$. Thus the electronic distribution becomes analytic
even in the ground state leading to the aforementioned exponential behavior. This result is precisely the same as found in BCS
superconductor in Ref.~[\onlinecite{fetter}]. Raising the temperature introduces a further exponential cutoff
($\sim\text{exp}[-2\pi|x|/\xi_1]$), just like it did for normal metal, that has clearly no observable effect as long as
$\xi_1\gg\xi_0$. The opposite relation on the other hand would lead essentially to the normal state.

Finite interchain coupling $t_b$ results in a quasi-one dimensional CDW structure. Its effect coincides exactly with that found
for normal metal because of the same factors of Bessel functions appearing in Eqs.~\eqref{bornmetal} and \eqref{borncdw}. For
details see the last paragraph of subsection A.

\subsection{Unconventional DW}
Now we finally turn our attention to UDW and for the same reasons pointed out in the case of CDW, we again consider the effect of
the relevant backscattering only. We begin with the zero temperature case. In the ground state a closed solution like the ones in
Eqs.~\eqref{bornmetal} and \eqref{borncdw} valid for all $|x|\gg v_F/D$ cannot be obtained. Nevertheless, we have just seen during
the calculation of the CDW response that within the coherence length the metallic result applies. This is because at such
distances only the high energy electron-hole pair excitations contribute to density oscillations, and far from the Fermi energy a
fully gapped CDW behaves the same as a normal metal. This latter statement is equally true for an UDW as well. Therefore we
conclude, that in quasi-one dimensional UDW ground state within the coherence length the density oscillations are the same as that
of a normal metal given by the zero temperature limit of Eq.~\eqref{bornmetal}. Note here, that in UDW due to the vanishing
average of the gap over the Fermi surface, the anomalous contribution to the total density due to the condensate is missing,
$n_1=0$. On the other hand, at large distances the small energy nodal excitations dominate the static charge response. To make the
picture whole, for $|x|\gg\xi_0$ asymptotic expansion to leading order yields

\begin{figure}[t]
\caption{\label{fig:friedel} The impurity induced electronic density is shown for a conventional CDW (solid) and an UDW
  (dashed-dotted) on the $m=0$ chain. It is calculated from Eqs.~\eqref{borncdw} and \eqref{bornudw}. For plotting
  $k_F\xi_0=k_F\xi=10$ were applied and $n^0(x,0)$ denotes the unperturbed density without impurity. The CDW response freezes out
  exponentially at large distances. The UDW contribution on the other hand is much larger which leads to the fact that the beat
  property is well observable too, in the present case with $k_F\lambda/2\approx\pi k_F\xi/2\approx15$.}
\end{figure}

\begin{equation}
\begin{split}
n(x,m)-n_0=&-n_0N_0U(\mathbf{Q})(-1)^m\cos^2\left(\frac{2x}{\xi}-\frac{\pi}{2}m\right)\frac{\cos(2k_Fx)}{2k_F\xi_0}
\frac{4z}{\left(2z^2+m^2\right)^2}\\
&+n_0N_0U(\mathbf{Q})(-1)^m\sin^2\left(\frac{2x}{\xi}-\frac{\pi}{2}m\right)\frac{\cos(2k_Fx+2\phi)}{2k_F\xi_0}
\frac{2m^2}{z\left(2z^2+m^2\right)^2},\label{bornudw}
\end{split}
\end{equation}
where $z=|x|/\xi_0$. Equation \eqref{bornudw} certifies our expectations about the power law decay $(\sim r^{-3})$ at large
distances. At this point we would like to call the attention to the following: The argumentation (about the smeared Fermi surface
leading to exponential decay) we applied in the CDW in subsection B is not applicable directly for an unconventional density wave,
though a finite order parameter exists here too. In fact the UDW gap is momentum dependent and possesses nodes. Around these nodes
the electronic distribution function $\langle a_{\mathbf{k}\sigma}^\dag
a_{\mathbf{k}\sigma}\rangle=2^{-1}(1-\xi(\mathbf{k})/\sqrt{\xi(\mathbf{k})^2+|\Delta(\mathbf{k})|^2})$ simplifies to the sharp
step function. At other portions of the Fermi surface however, where $|\Delta(\mathbf{k})|$ is close to maximum, the distribution
function is smeared out much the same way as in normal CDW. Nevertheless, due to nodal quasiparticles the distribution function
remains non-analytical in momentum space leading to the observed long-range power law decay.

At finite temperature charge response is affected only at distances $|x|>>\xi_1$, where the factor $\text{exp}[-2\pi|x|/\xi_1]$
becomes dominant. But as long as $\beta\Delta(T)=\xi_1/\xi_0\gg1$, the power law behavior can in principle be observed in the
range $\xi_0\ll|x|\ll\xi_1$. In Fig.~\ref{fig:friedel} the density oscillations along the $m=0$ chain are compared in a
conventional CDW and an UDW. Beyond the coherence length the CDW response is hardly observable while it is considerably larger in
the UDW. Furthermore the aforementioned beat due to finite interchain coupling can be clearly seen in the latter case too. This
complex behavior signals the presence of the nodal density wave, and seems to be more accessible in experiments than the detection
of an exponential decay in normal CDW. In the latter case of course, clear CDW background is present too, that is certainly
measurable by other means. In UDW candidates however, measuring Friedel oscillations for example in STM measurements might serve
as a useful tool in identifying the low temperature phase and to reveal hidden order.

\section{Thermodynamics}
The effect of the single non-magnetic impurity on the thermodynamics can be derived from Eq.~\eqref{domega}, which gives the
change in the grand canonical potential in the presence of interaction. This equation will be utilized in this section in order to
calculate the entropy and specific heat contribution of dilute impurities. According to Eq.~\eqref{domega}, we need the exact
Green's matrix at the impurity site and at $\tau=-0$ imaginary time. The Matsubara summation is transformed, as usual by deforming
the contour, into a frequency integration along the real axis. This form now involves the spectral function at the impurity site,
and we get
\begin{equation}
G_{\alpha\beta}(0,0;\tau<0)=\frac{1}{4}\int_{-\infty}^\infty d\omega\,f(\omega)e^{-\omega\tau}[A_+(\omega)+\alpha\beta
  A_-(\omega)],\qquad\alpha,\beta=R(=+1),\,L(=-1),\label{summedgreen}
\end{equation}
where $f(\omega)$ is the Fermi function, $A_+(\omega)=A(\mathbf{r}=0,\omega)$, and $A_-$ has the same functional form as $A_+$,
but instead of $U_+$ it involves $U_-$, that is
\begin{equation}
A_\pm(\omega)=-\frac2\pi\frac{g_2}{(1-U_\pm g_1)^2+(U_\pm g_2)^2}+ 2g_1(\omega_\pm)^2|\partial_\omega
g_1(\omega_\pm)|^{-1}\delta(\omega-\omega_\pm).\label{apm}
\end{equation}

\begin{figure}[t]
\caption{\label{fig:ldosimp}The local density of states is shown at the impurity site versus energy and the impurity strength,
  where $U_+=U(0)+U(\mathbf{Q})$.}
\end{figure}

\noindent Here $g_{1,2}$ stand for the real and imaginary parts of the local zeroth order Green's function given by
Eqs.~\eqref{g1} and \eqref{g2} in Appendix A. The LDOS right at the impurity site $A(\mathbf{r}=0,\omega)$ is shown in
Fig.~\ref{fig:ldosimp}. In addition, $\omega_\pm$ denote the binding energies of the high energy bound states outside the band
determined from $1-U_\pm g_1(\omega_\pm)=0$. This is nothing else but Eq.~\eqref{center} for frequencies where $g_2=0$, i.e.
outside the band. It provides us those poles of the $T$-matrix that are located on the real frequency axis. With the aid of
Eqs.~\eqref{g1metal} and \eqref{g2metal} the solution reads as $\omega_\pm=D\coth(1/N_0U_\pm)$. These states are indeed very far
from the Fermi level, their contribution to the low temperature behavior is obviously negligible. Doing so and keeping only the
first term in Eq.~\eqref{apm}, we are able to perform the coupling constant integral in Eq.~\eqref{domega} analytically, and we
end up with
\begin{equation}
\delta\Omega=-\frac2\pi\int_{-\infty}^\infty d\omega\,f(\omega)
\left[ \arctan\left(\frac{(g_1^2+g_2^2)U_+-g_1}{g_2}\right)+\arctan\left(\frac{(g_1^2+g_2^2)U_- -g_1}{g_2}\right)
+2\arctan\left(\frac{g_1}{g_2}\right)\right].
\end{equation}
We note at this point that the temperature enters this expression not only through the Fermi function, but also through the UDW
order parameter $\Delta(T)$ appearing in $g_{1,2}$. Nevertheless, the low temperature correction to $\Delta(0)$ is small, it is of
the order of $T^3$,\cite{balazs-sdw} and thus cannot affect the expected $T$ linear entropy and specific
heat.\cite{balazs-born,balazs-unitary,nca-impurity} A Bethe-Sommerfeld series expansion at low $T$ finally yields the specific
heat contribution as
\begin{equation}
\delta C_V=\frac{2\pi^2}{3}\frac{1}{V}\frac{N_0U(0)}{\Delta(0)}T,\label{heat}
\end{equation}
where $1/V$ is the impurity concentration in this single impurity problem. It should be replaced by $n_i=N_i/V$ in the extreme
dilute limit, where $N_i$ is the number of impurity atoms in the sample. Equation \eqref{heat} shows that the $T$ linear
contribution is due to forward scattering. Furthermore, $n_iN_0U(0)/\Delta$ plays the role of the residual density of states at
the Fermi level in an impure UDW. These findings are in accordance with Ref.~[\onlinecite{nca-impurity}].

\section{Conclusion}
We have studied the effect of a single non-magnetic impurity in a quasi-one dimensional unconventional density wave. In
particular, its effect on the local electronic states and on the Friedel oscillations were explored in great detail. The potential
scattering was treated within the $T$-matrix approach, with the allowance for different forward and backward scattering. In this
respect, we extended the widely applied strictly point-like impurity picture. The impurity induced local density of states becomes
asymmetric with respect to the Fermi energy, signalling the violation of particle-hole symmetry. We found a double peaked
quasiparticle resonance in the subgap, calculated the energy and lifetime of the virtual states, and determined the scanning
tunneling microscopy image along the neighboring chains. The double peaked nature of the virtual states stems from the
generalization for different scattering amplitudes. Indeed, for equal forward and backscattering corresponding to a delta
potential, we obtain a single impurity state in accordance with that found in quasi-two dimensional dDW. The energies and decay
rates are found the same as in dDW or dSC, which is not surprising at all, as these systems all have a quasiparticle density of
states that is linear in energy around the Fermi level arising from nodal excitations. The electronic states around the impurity
can be studied experimentally with STM spectroscopy. This diagnostic tool is able to resolve both the energy and spatial
dependence of the local density of states by directly measuring the tunneling conductance. To this end, we calculated the expected
STM image of an UDW as a function of energy and position.

An external perturbation like an added impurity, or a constraint on the electronic wave functions in the form of boundary
conditions, are known to cause Friedel oscillations in the charge distribution. In pure UDW no modulation of either charge or spin
is present. Therefore, robust Friedel oscillations were expected to show up below the density wave coherence length. We found
indeed, that in this length scale the density oscillations were those of a normal metal. This is to be contrasted with the CDW
result, where the impurity induced oscillations are superimposed on the usual CDW background. On the other hand, contrary to the
exponential decay of fully gapped CDW, beyond the coherence length power law behavior was expected and found. It is due to the
fact, that in UDW nodal excitations (electron-hole pair) are available at arbitrarily small energy, thus density oscillations can
build up. This algebraic behavior at large distances signals the presence of the UDW condensate and could be more accessible in
experiments than the exponential decay of normal CDW. In addition to that, the possibility of experimental detection is further
supported by the fact that an UDW does not exhibit static charge density wave background that could overwhelm the impurity
contribution.

At last, we calculated the change in the grand canonical potential, the entropy and specific heat contribution of the scalar
impurity embedded in the UDW host. The calculation was done to infinite order in the interaction. At sufficiently low temperature
forward scattering produces metallic $T$-linear behavior, because it breaks particle-hole symmetry and causes finite residual
density of states at the Fermi energy.

\begin{acknowledgments}
This work was supported by the Hungarian National Research Fund under grant numbers OTKA NDF45172, NI70594, T046269, TS049881.
\end{acknowledgments}

\appendix
\section{}
In this Appendix we calculate the zeroth order retarded Green's matrix $G^0_{\alpha\beta}(\mathbf{r,r'};\omega)$ of an
unconventional density wave in real space. With the aid of the definition in Eq.~\eqref{definition}, and using the form of left-,
and right-moving fields, see Eq.~\eqref{fieldoperator}, one obtains
\begin{equation}
G^0_{\alpha\beta}(\mathbf{r,r'};\omega)=e^{i(\mathbf{Q}/2)[(\alpha-1)\mathbf{r}-(\beta-1)\mathbf{r'}]
+i(\phi_\alpha-\phi_\beta)}\hat G^0_{\alpha\beta}(\mathbf{r-r'},\omega),\qquad\alpha,\beta=R(=+1),\,L(=-1),\label{retarded}
\end{equation}
where $\phi=\phi_R-\phi_L$ is the phase of the order parameter. As we detached this complex phase factor from the order parameter,
it is now purely real: $\Delta(\mathbf{k})=\Delta\sin(bk_y)$. Furthermore, the homogeneous part reads as
\begin{equation}
\hat G^0(\mathbf{r},\omega)=\frac{1}{V}\sideset{}{'}\sum_\mathbf{k}e^{i\mathbf{kr}}
\frac{\omega+\xi(\mathbf{k})\rho_3+\Delta(\mathbf{k})\rho_1}{\omega^2-\xi(\mathbf{k})^2-\Delta(\mathbf{k})^2+\text{sign}(\omega)i0},
\end{equation}
with $\rho_i$ the Pauli matrices, and the prime on the momentum integration again denotes that the sum is cut off as
$|k_x-k_F|<k_c$, where $v_Fk_c\equiv D$ is half the bandwidth. We note that in a strictly one-dimensional (no perpendicular
coupling between parallel chains, $t_b=0$) conventional density wave with constant gap, one only needs to perform the substitution
$\sin(bk_y)\to1$ in order to get the desired results. In particular, for the diagonal components one finds
\begin{equation}
\begin{split}
\hat G^0_{\alpha\alpha} & (x,\omega)_\text{conv}=-\frac{1}{4\pi v_F}e^{x\left(ik_F+\frac{\sqrt{\Delta^2-\omega^2}}{v_F}\right)}
\left(\frac{i\omega}{\sqrt{\Delta^2-\omega^2}}+\alpha\right)\\
&\times\left[\text{E}_1\left(\frac{ix}{v_F}(D-i\sqrt{\Delta^2-\omega^2})\right)
-\text{E}_1\left(\frac{ix}{v_F}(-D-i\sqrt{\Delta^2-\omega^2})\right)
-2i\pi\Theta(-x)\Theta(D'-|\omega|)\right]\\
&+\frac{1}{4\pi v_F}e^{x\left(ik_F-\frac{\sqrt{\Delta^2-\omega^2}}{v_F}\right)}
\left(\frac{i\omega}{\sqrt{\Delta^2-\omega^2}}-\alpha\right)\\
&\times\left[\text{E}_1\left(\frac{ix}{v_F}(D+i\sqrt{\Delta^2-\omega^2})\right)
-\text{E}_1\left(\frac{ix}{v_F}(-D+i\sqrt{\Delta^2-\omega^2})\right)
+2i\pi\Theta(x)\Theta(D'-|\omega|)\right],\label{g0rr}
\end{split}
\end{equation}
and the offdiagonal components read as
\begin{equation}
\begin{split}
\hat G^0_{\alpha,-\alpha}(x,\omega)_\text{conv}=&-\frac{1}{4\pi v_F}\frac{i\Delta}{\sqrt{\Delta^2-\omega^2}}
e^{x\left(ik_F+\frac{\sqrt{\Delta^2-\omega^2}}{v_F}\right)}\\
&\times\left[\text{E}_1\left(\frac{ix}{v_F}(D-i\sqrt{\Delta^2-\omega^2})\right)
-\text{E}_1\left(\frac{ix}{v_F}(-D-i\sqrt{\Delta^2-\omega^2})\right) -2i\pi\Theta(-x)\Theta(D'-|\omega|)\right]\\
&+\frac{1}{4\pi v_F}\frac{i\Delta}{\sqrt{\Delta^2-\omega^2}}e^{x\left(ik_F-\frac{\sqrt{\Delta^2-\omega^2}}{v_F}\right)}\\
&\times\left[\text{E}_1\left(\frac{ix}{v_F}(D+i\sqrt{\Delta^2-\omega^2})\right)
-\text{E}_1\left(\frac{ix}{v_F}(-D+i\sqrt{\Delta^2-\omega^2})\right) +2i\pi\Theta(x)\Theta(D'-|\omega|)\right].\label{g0rl}
\end{split}
\end{equation}
Here $\text{E}_1(z)$ is the exponential integral, $\omega$ means $\omega+i0$ under the squareroots, and
$D'=\sqrt{D^2+\Delta^2}$. These formulas reproduce those of Ref.~[\onlinecite{tutto}] in the limit $D\to\infty$. Now it is
convenient to point out, that the corresponding results for a strictly one-dimensional normal metal can be obtained easily with
the substitution $\Delta=0$, and one obtains the usual diagonal and translational invariant solution.

After introducing the strictly one-dimensional results, we now turn our attention to quasi-one dimensional systems with finite
interchain coupling, $t_b\ne0$. These include the normal metal, a conventional DW, or UDW. We are primarily interested in the
latter system, because this is the one that exists only in dimensions greater than one. However, as the UDW formalism is the most
general with its momentum dependent gap, it incorporates the results for each case. Quasi-one dimensionality means that there is
at least one more (perpendicular) direction in real space, where the model is discrete rather than being continuous as it is in
the chain direction $x$. In our notation this additional dimension was chosen to be the $y$ direction, where the order parameter
varies. In this respect, the position argument of the Green's function is $\mathbf{r}=(x,mb,0)$, where $b$ is the corresponding
lattice constant and $m$ is an integer indexing parallel chains. With all this, the analogous results for the bare Green's
function in UDW are obtained from Eqs.~\eqref{g0rr} and \eqref{g0rl} by performing first the substitutions
$\Delta\to\Delta\sin(bk_y)$, $\pm D\to\pm D-2t_b\cos(bk_y)$, and then integrating over the Fermi surface
\begin{equation}
\hat G^0_{\alpha\beta}(x,m;\omega)=\int_0^{2\pi}\frac{d(bk_y)}{2\pi bc}e^{i\left(bk_ym+\frac{2x}{\xi}\cos(bk_y)\right)}
\hat G^0_{\alpha\beta}(x,\omega)_\text{conv},\label{udwgreen}
\end{equation}
where $\xi=v_F/t_b$. The integration in Eq.~\eqref{udwgreen} can be performed for a normal metal and a conventional density wave
in the $|x|\gg v_F/D$ limit, that is for distances much larger than the atomic lengthscale. It leads to the appearance of
$J_m(2x/\xi)$, the Bessel function of the first kind. For UDW, since the integration cannot be carried out, the Bessel function
does not show up explicitly. Nevertheless its properties are coded in the integral and we will indeed encounter some of them
during the calculation of the spectral function and Friedel oscillations.

We end this Appendix with the presentation of the local Green's function in UDW. At the impurity site $x=0$, $m=0$ the averaging
over the Fermi surface can be done analytically and one finds $\hat G_{\alpha\beta}^0=\delta_{\alpha\beta}(g_1+ig_2)$, where
\begin{equation}
g_1(\omega)\equiv\text{Re}\hat G^0_{RR}(0,0;\omega)=
\begin{cases}
-N_0\frac{\omega}{\Delta}F\left(\frac{\sqrt{D^2\Delta^2-\omega^2(D^2+\Delta^2-\omega^2)}}{D\sqrt{\Delta^2-\omega^2}},
\frac{\sqrt{\Delta^2-\omega^2}}{\Delta}\right) &\qquad 0<|\omega|<\Delta,\\[8pt]
N_0F\left(\frac{\omega}{D},\frac{\sqrt{\omega^2-\Delta^2}}{\omega}\right) &\qquad \Delta<|\omega|<D,\\[8pt]
N_0\text{sign}(\omega) K\left(\frac{\sqrt{\omega^2-\Delta^2}}{\omega}\right) &\qquad D<|\omega|<\sqrt{D^2+\Delta^2},\\[8pt]
N_0\text{sign}(\omega) F\left(\frac{D}{\sqrt{\omega^2-\Delta^2}},\frac{\sqrt{\omega^2-\Delta^2}}{\omega}\right) &\qquad
\sqrt{D^2+\Delta^2}<|\omega|,
\end{cases}\label{g1}
\end{equation}
\begin{equation}
g_2(\omega)\equiv\text{Im}\hat G^0_{RR}(0,0;\omega)=
\begin{cases}
-N_0\frac{|\omega|}{\Delta}K\left(\frac{\omega}{\Delta}\right)&\qquad 0<|\omega|<\Delta,\\[8pt]
-N_0K\left(\frac{\Delta}{\omega}\right) &\qquad \Delta<|\omega|<D,\\[8pt]
-N_0\left[K\left(\frac{\Delta}{\omega}\right)-F\left(\frac{\sqrt{\omega^2-D^2}}{\Delta},\frac{\Delta}{\omega}\right)\right]
&\qquad D<|\omega|<\sqrt{D^2+\Delta^2},\\[8pt]
0 &\qquad \sqrt{D^2+\Delta^2}<|\omega|.
\end{cases}\label{g2}
\end{equation}
Here $N_0=1/(\pi v_Fbc)$ is the normal state density of states at the Fermi level (per unit volume and for one spin direction),
and $K(k)$ and $F(z,k)$ are, respectively, the complete and incomplete elliptic integrals of the first kind. During the
calculation of Eqs.~\eqref{g1} and \eqref{g2} a correction of order $\mathcal{O}(t_b/D)$ was neglected as it only affects the
frequency dependence at high energies around the band edges, and is therefore irrelevant close to the Fermi level. In addition to
that, in this case the bandwidth can also be taken to infinity and the above formulas simplify to
\begin{align}
g_1&=-N_0\Theta(\Delta-|\omega|)\frac{\omega}{\Delta}K\left(\sqrt{1-\left(\frac{\omega}{\Delta}\right)^2}\right),\label{g1new}\\
g_2&=-N_0\left(\Theta(\Delta-|\omega|)\frac{|\omega|}{\Delta}K\left(\frac{\omega}{\Delta}\right)+
\Theta(|\omega|-\Delta)K\left(\frac{\Delta}{\omega}\right)\right).\label{g2new}
\end{align}
On the other hand, if we are interested in the high energy behavior close to band edges, $\Delta$ can be neglected compared to $D$
and Eqs.~\eqref{g1} and \eqref{g2} simplify to
\begin{align}
g_1&=\frac{N_0}{2}\ln\left|\frac{\omega+D}{\omega-D}\right|,\label{g1metal}\\
g_2&=-\frac{\pi N_0}{2}\Theta(D-|\omega|).\label{g2metal}
\end{align}

\bibliography{friedel}

\begin{thebibliography}{38}
\expandafter\ifx\csname natexlab\endcsname\relax\def\natexlab#1{#1}\fi
\expandafter\ifx\csname bibnamefont\endcsname\relax
  \def\bibnamefont#1{#1}\fi
\expandafter\ifx\csname bibfnamefont\endcsname\relax
  \def\bibfnamefont#1{#1}\fi
\expandafter\ifx\csname citenamefont\endcsname\relax
  \def\citenamefont#1{#1}\fi
\expandafter\ifx\csname url\endcsname\relax
  \def\url#1{\texttt{#1}}\fi
\expandafter\ifx\csname urlprefix\endcsname\relax\def\urlprefix{URL }\fi
\providecommand{\bibinfo}[2]{#2}
\providecommand{\eprint}[2][]{\url{#2}}

\bibitem[{\citenamefont{Morr}(2002)}]{morr}
\bibinfo{author}{\bibfnamefont{D.~K.} \bibnamefont{Morr}},
  \bibinfo{journal}{Phys. Rev. Lett.} \textbf{\bibinfo{volume}{89}},
  \bibinfo{pages}{106401} (\bibinfo{year}{2002}).

\bibitem[{\citenamefont{Zhu et~al.}(2001)\citenamefont{Zhu, Kim, Ting, and
  Carbotte}}]{jian}
\bibinfo{author}{\bibfnamefont{J.~X.} \bibnamefont{Zhu}},
  \bibinfo{author}{\bibfnamefont{W.}~\bibnamefont{Kim}},
  \bibinfo{author}{\bibfnamefont{C.~S.} \bibnamefont{Ting}}, \bibnamefont{and}
  \bibinfo{author}{\bibfnamefont{J.~P.} \bibnamefont{Carbotte}},
  \bibinfo{journal}{Phys. Rev. Lett.} \textbf{\bibinfo{volume}{87}},
  \bibinfo{pages}{197001} (\bibinfo{year}{2001}).

\bibitem[{\citenamefont{Wang}(2002)}]{quiang}
\bibinfo{author}{\bibfnamefont{Q.~H.} \bibnamefont{Wang}},
  \bibinfo{journal}{Phys. Rev. Lett.} \textbf{\bibinfo{volume}{88}},
  \bibinfo{pages}{057002} (\bibinfo{year}{2002}).

\bibitem[{\citenamefont{Salkola et~al.}(1996)\citenamefont{Salkola, Balatsky,
  and Scalapino}}]{salkola}
\bibinfo{author}{\bibfnamefont{M.~I.} \bibnamefont{Salkola}},
  \bibinfo{author}{\bibfnamefont{A.~V.} \bibnamefont{Balatsky}},
  \bibnamefont{and} \bibinfo{author}{\bibfnamefont{D.~J.}
  \bibnamefont{Scalapino}}, \bibinfo{journal}{Phys. Rev. Lett.}
  \textbf{\bibinfo{volume}{77}}, \bibinfo{pages}{1841} (\bibinfo{year}{1996}).

\bibitem[{\citenamefont{Balatsky et~al.}(1995)\citenamefont{Balatsky, Salkola,
  and Rosengren}}]{balatsky}
\bibinfo{author}{\bibfnamefont{A.~V.} \bibnamefont{Balatsky}},
  \bibinfo{author}{\bibfnamefont{M.~I.} \bibnamefont{Salkola}},
  \bibnamefont{and}
  \bibinfo{author}{\bibfnamefont{A.}~\bibnamefont{Rosengren}},
  \bibinfo{journal}{Phys. Rev. B} \textbf{\bibinfo{volume}{51}},
  \bibinfo{pages}{15547} (\bibinfo{year}{1995}).

\bibitem[{\citenamefont{Balatsky et~al.}(2006)\citenamefont{Balatsky, Vekhter,
  and Zhu}}]{balatsky-review}
\bibinfo{author}{\bibfnamefont{A.~V.} \bibnamefont{Balatsky}},
  \bibinfo{author}{\bibfnamefont{I.}~\bibnamefont{Vekhter}}, \bibnamefont{and}
  \bibinfo{author}{\bibfnamefont{J.~X.} \bibnamefont{Zhu}},
  \bibinfo{journal}{Rev. Mod. Phys.} \textbf{\bibinfo{volume}{78}},
  \bibinfo{pages}{373} (\bibinfo{year}{2006}).

\bibitem[{\citenamefont{Pan et~al.}(2000)\citenamefont{Pan, Hudson, Lang,
  Eisaki, Uchida, and Davis}}]{pan}
\bibinfo{author}{\bibfnamefont{S.~H.} \bibnamefont{Pan}},
  \bibinfo{author}{\bibfnamefont{E.~W.} \bibnamefont{Hudson}},
  \bibinfo{author}{\bibfnamefont{K.~M.} \bibnamefont{Lang}},
  \bibinfo{author}{\bibfnamefont{H.}~\bibnamefont{Eisaki}},
  \bibinfo{author}{\bibfnamefont{S.}~\bibnamefont{Uchida}}, \bibnamefont{and}
  \bibinfo{author}{\bibfnamefont{J.~C.} \bibnamefont{Davis}},
  \bibinfo{journal}{Nature (London)} \textbf{\bibinfo{volume}{403}},
  \bibinfo{pages}{746} (\bibinfo{year}{2000}).

\bibitem[{\citenamefont{Hudson et~al.}(1999)\citenamefont{Hudson, Pan, Gupta,
  Ng, and Davis}}]{hudson}
\bibinfo{author}{\bibfnamefont{E.~W.} \bibnamefont{Hudson}},
  \bibinfo{author}{\bibfnamefont{S.~H.} \bibnamefont{Pan}},
  \bibinfo{author}{\bibfnamefont{A.~K.} \bibnamefont{Gupta}},
  \bibinfo{author}{\bibfnamefont{K.~W.} \bibnamefont{Ng}}, \bibnamefont{and}
  \bibinfo{author}{\bibfnamefont{J.~C.} \bibnamefont{Davis}},
  \bibinfo{journal}{Science} \textbf{\bibinfo{volume}{285}},
  \bibinfo{pages}{88} (\bibinfo{year}{1999}).

\bibitem[{\citenamefont{Yazdani et~al.}(1999)\citenamefont{Yazdani, Howald,
  Lutz, Kapitulnik, and Eigler}}]{yazdani}
\bibinfo{author}{\bibfnamefont{A.}~\bibnamefont{Yazdani}},
  \bibinfo{author}{\bibfnamefont{C.~M.} \bibnamefont{Howald}},
  \bibinfo{author}{\bibfnamefont{C.~P.} \bibnamefont{Lutz}},
  \bibinfo{author}{\bibfnamefont{A.}~\bibnamefont{Kapitulnik}},
  \bibnamefont{and} \bibinfo{author}{\bibfnamefont{D.~M.}
  \bibnamefont{Eigler}}, \bibinfo{journal}{Phys. Rev. Lett.}
  \textbf{\bibinfo{volume}{83}}, \bibinfo{pages}{176} (\bibinfo{year}{1999}).

\bibitem[{\citenamefont{Nayak}(2000)}]{nayak-solo}
\bibinfo{author}{\bibfnamefont{C.}~\bibnamefont{Nayak}},
  \bibinfo{journal}{Phys. Rev. B} \textbf{\bibinfo{volume}{62}},
  \bibinfo{pages}{4880} (\bibinfo{year}{2000}).

\bibitem[{\citenamefont{Chakravarty et~al.}(2001)\citenamefont{Chakravarty,
  Laughlin, Morr, and Nayak}}]{laughlin}
\bibinfo{author}{\bibfnamefont{S.}~\bibnamefont{Chakravarty}},
  \bibinfo{author}{\bibfnamefont{R.~B.} \bibnamefont{Laughlin}},
  \bibinfo{author}{\bibfnamefont{D.~K.} \bibnamefont{Morr}}, \bibnamefont{and}
  \bibinfo{author}{\bibfnamefont{C.}~\bibnamefont{Nayak}},
  \bibinfo{journal}{Phys. Rev. B} \textbf{\bibinfo{volume}{63}},
  \bibinfo{pages}{094503} (\bibinfo{year}{2001}).

\bibitem[{\citenamefont{Emery and Kivelson}(1995)}]{emery}
\bibinfo{author}{\bibfnamefont{V.~J.} \bibnamefont{Emery}} \bibnamefont{and}
  \bibinfo{author}{\bibfnamefont{S.~A.} \bibnamefont{Kivelson}},
  \bibinfo{journal}{Nature (London)} \textbf{\bibinfo{volume}{374}},
  \bibinfo{pages}{434} (\bibinfo{year}{1995}).

\bibitem[{\citenamefont{Renner et~al.}(1998)\citenamefont{Renner, Revaz,
  Genoud, Kadowaki, and Fischer}}]{renner}
\bibinfo{author}{\bibfnamefont{C.}~\bibnamefont{Renner}},
  \bibinfo{author}{\bibfnamefont{B.}~\bibnamefont{Revaz}},
  \bibinfo{author}{\bibfnamefont{J.~Y.} \bibnamefont{Genoud}},
  \bibinfo{author}{\bibfnamefont{K.}~\bibnamefont{Kadowaki}}, \bibnamefont{and}
  \bibinfo{author}{\bibfnamefont{O.}~\bibnamefont{Fischer}},
  \bibinfo{journal}{Phys. Rev. Lett.} \textbf{\bibinfo{volume}{80}},
  \bibinfo{pages}{149} (\bibinfo{year}{1998}).

\bibitem[{\citenamefont{Kruis et~al.}(2001)\citenamefont{Kruis, Martin, and
  Balatsky}}]{kruis}
\bibinfo{author}{\bibfnamefont{H.~V.} \bibnamefont{Kruis}},
  \bibinfo{author}{\bibfnamefont{I.}~\bibnamefont{Martin}}, \bibnamefont{and}
  \bibinfo{author}{\bibfnamefont{A.~V.} \bibnamefont{Balatsky}},
  \bibinfo{journal}{Phys. Rev. B} \textbf{\bibinfo{volume}{64}},
  \bibinfo{pages}{054501} (\bibinfo{year}{2001}).

\bibitem[{\citenamefont{Shiba}(1968)}]{shiba}
\bibinfo{author}{\bibfnamefont{H.}~\bibnamefont{Shiba}},
  \bibinfo{journal}{Prog. Theor. Phys.} \textbf{\bibinfo{volume}{40}},
  \bibinfo{pages}{435} (\bibinfo{year}{1968}).

\bibitem[{\citenamefont{Machida and Shibata}(1972)}]{machida}
\bibinfo{author}{\bibfnamefont{K.}~\bibnamefont{Machida}} \bibnamefont{and}
  \bibinfo{author}{\bibfnamefont{F.}~\bibnamefont{Shibata}},
  \bibinfo{journal}{Prog. Theor. Phys.} \textbf{\bibinfo{volume}{47}},
  \bibinfo{pages}{1817} (\bibinfo{year}{1972}).

\bibitem[{\citenamefont{Hotta}(1993)}]{hotta}
\bibinfo{author}{\bibfnamefont{T.}~\bibnamefont{Hotta}}, \bibinfo{journal}{J.
  Phys. Soc. Jpn.} \textbf{\bibinfo{volume}{62}}, \bibinfo{pages}{274}
  (\bibinfo{year}{1993}).

\bibitem[{\citenamefont{T\"utt\H{o} and Zawadowski}(1985)}]{tutto}
\bibinfo{author}{\bibfnamefont{I.}~\bibnamefont{T\"utt\H{o}}} \bibnamefont{and}
  \bibinfo{author}{\bibfnamefont{A.}~\bibnamefont{Zawadowski}},
  \bibinfo{journal}{Phys. Rev. B} \textbf{\bibinfo{volume}{32}},
  \bibinfo{pages}{2449} (\bibinfo{year}{1985}).

\bibitem[{\citenamefont{Cheng}(1987)}]{cheng}
\bibinfo{author}{\bibfnamefont{L.~J.} \bibnamefont{Cheng}},
  \bibinfo{journal}{J. Phys. C.} \textbf{\bibinfo{volume}{20}},
  \bibinfo{pages}{4917} (\bibinfo{year}{1987}).

\bibitem[{\citenamefont{Zawadowski and T\"utt\H{o}}(1989)}]{zawatutto}
\bibinfo{author}{\bibfnamefont{A.}~\bibnamefont{Zawadowski}} \bibnamefont{and}
  \bibinfo{author}{\bibfnamefont{I.}~\bibnamefont{T\"utt\H{o}}},
  \bibinfo{journal}{Synth. Metals} \textbf{\bibinfo{volume}{29}},
  \bibinfo{pages}{469} (\bibinfo{year}{1989}).

\bibitem[{\citenamefont{Hansen and Baeriswyl}(1986)}]{hansen}
\bibinfo{author}{\bibfnamefont{L.~K.} \bibnamefont{Hansen}} \bibnamefont{and}
  \bibinfo{author}{\bibfnamefont{D.}~\bibnamefont{Baeriswyl}},
  \bibinfo{journal}{J. Phys. C: Solid State Phys.}
  \textbf{\bibinfo{volume}{19}}, \bibinfo{pages}{5615} (\bibinfo{year}{1986}).

\bibitem[{\citenamefont{D\'ora et~al.}(2004)\citenamefont{D\'ora, Maki, and
  Virosztek}}]{balazs-modern}
\bibinfo{author}{\bibfnamefont{B.}~\bibnamefont{D\'ora}},
  \bibinfo{author}{\bibfnamefont{K.}~\bibnamefont{Maki}}, \bibnamefont{and}
  \bibinfo{author}{\bibfnamefont{A.}~\bibnamefont{Virosztek}},
  \bibinfo{journal}{Mod. Phys. Lett. B} \textbf{\bibinfo{volume}{18}},
  \bibinfo{pages}{327} (\bibinfo{year}{2004}).

\bibitem[{\citenamefont{D\'ora and Virosztek}(2001)}]{balazs-sdw}
\bibinfo{author}{\bibfnamefont{B.}~\bibnamefont{D\'ora}} \bibnamefont{and}
  \bibinfo{author}{\bibfnamefont{A.}~\bibnamefont{Virosztek}},
  \bibinfo{journal}{Eur. Phys. J. B} \textbf{\bibinfo{volume}{22}},
  \bibinfo{pages}{167} (\bibinfo{year}{2001}).

\bibitem[{\citenamefont{Andres et~al.}(2001)\citenamefont{Andres, Kartsovnik,
  Biberacher, Weiss, Balthes, M\"uller, and Kushch}}]{andres}
\bibinfo{author}{\bibfnamefont{D.}~\bibnamefont{Andres}},
  \bibinfo{author}{\bibfnamefont{M.~V.} \bibnamefont{Kartsovnik}},
  \bibinfo{author}{\bibfnamefont{W.}~\bibnamefont{Biberacher}},
  \bibinfo{author}{\bibfnamefont{H.}~\bibnamefont{Weiss}},
  \bibinfo{author}{\bibfnamefont{E.}~\bibnamefont{Balthes}},
  \bibinfo{author}{\bibfnamefont{H.}~\bibnamefont{M\"uller}}, \bibnamefont{and}
  \bibinfo{author}{\bibfnamefont{N.}~\bibnamefont{Kushch}},
  \bibinfo{journal}{Phys. Rev. B} \textbf{\bibinfo{volume}{64}},
  \bibinfo{pages}{161104(R)} (\bibinfo{year}{2001}).

\bibitem[{\citenamefont{Mori et~al.}(1990)\citenamefont{Mori, Tanaka, Oshima,
  Saito, Mori, Maruyama, and Inokuchi}}]{mori}
\bibinfo{author}{\bibfnamefont{H.}~\bibnamefont{Mori}},
  \bibinfo{author}{\bibfnamefont{S.}~\bibnamefont{Tanaka}},
  \bibinfo{author}{\bibfnamefont{M.}~\bibnamefont{Oshima}},
  \bibinfo{author}{\bibfnamefont{G.}~\bibnamefont{Saito}},
  \bibinfo{author}{\bibfnamefont{T.}~\bibnamefont{Mori}},
  \bibinfo{author}{\bibfnamefont{Y.}~\bibnamefont{Maruyama}}, \bibnamefont{and}
  \bibinfo{author}{\bibfnamefont{H.}~\bibnamefont{Inokuchi}},
  \bibinfo{journal}{Bull. Chem. Soc. Jpn.} \textbf{\bibinfo{volume}{63}},
  \bibinfo{pages}{2138} (\bibinfo{year}{1990}).

\bibitem[{\citenamefont{Kartsovnik et~al.}(1995)\citenamefont{Kartsovnik,
  Kovalev, Laukhin, Schegolev, Ito, Ishiguro, Kushch, Mori, and
  Saito}}]{kartsovnik}
\bibinfo{author}{\bibfnamefont{M.~V.} \bibnamefont{Kartsovnik}},
  \bibinfo{author}{\bibfnamefont{A.~E.} \bibnamefont{Kovalev}},
  \bibinfo{author}{\bibfnamefont{V.~N.} \bibnamefont{Laukhin}},
  \bibinfo{author}{\bibfnamefont{I.~F.} \bibnamefont{Schegolev}},
  \bibinfo{author}{\bibfnamefont{H.}~\bibnamefont{Ito}},
  \bibinfo{author}{\bibfnamefont{T.}~\bibnamefont{Ishiguro}},
  \bibinfo{author}{\bibfnamefont{N.~D.} \bibnamefont{Kushch}},
  \bibinfo{author}{\bibfnamefont{H.}~\bibnamefont{Mori}}, \bibnamefont{and}
  \bibinfo{author}{\bibfnamefont{G.}~\bibnamefont{Saito}},
  \bibinfo{journal}{Synth. Metals} \textbf{\bibinfo{volume}{70}},
  \bibinfo{pages}{811} (\bibinfo{year}{1995}).

\bibitem[{\citenamefont{Basleti\'c et~al.}(2001)\citenamefont{Basleti\'c,
  Korin-Hamzi\'c, Kartsovnik, and M\"uller}}]{basletic}
\bibinfo{author}{\bibfnamefont{M.}~\bibnamefont{Basleti\'c}},
  \bibinfo{author}{\bibfnamefont{B.}~\bibnamefont{Korin-Hamzi\'c}},
  \bibinfo{author}{\bibfnamefont{M.~V.} \bibnamefont{Kartsovnik}},
  \bibnamefont{and} \bibinfo{author}{\bibfnamefont{H.}~\bibnamefont{M\"uller}},
  \bibinfo{journal}{Synth. Metals} \textbf{\bibinfo{volume}{120}},
  \bibinfo{pages}{1021} (\bibinfo{year}{2001}).

\bibitem[{\citenamefont{Fujita et~al.}(2001)\citenamefont{Fujita, Sasaki,
  Yoneyama, Kobayashi, and Fukase}}]{fujita}
\bibinfo{author}{\bibfnamefont{T.}~\bibnamefont{Fujita}},
  \bibinfo{author}{\bibfnamefont{T.}~\bibnamefont{Sasaki}},
  \bibinfo{author}{\bibfnamefont{N.}~\bibnamefont{Yoneyama}},
  \bibinfo{author}{\bibfnamefont{N.}~\bibnamefont{Kobayashi}},
  \bibnamefont{and} \bibinfo{author}{\bibfnamefont{T.}~\bibnamefont{Fukase}},
  \bibinfo{journal}{Synth. Metals} \textbf{\bibinfo{volume}{120}},
  \bibinfo{pages}{1077} (\bibinfo{year}{2001}).

\bibitem[{\citenamefont{Leylekian et~al.}(2003)\citenamefont{Leylekian, Ravy,
  and Pouget}}]{pouget}
\bibinfo{author}{\bibfnamefont{P.~F.} \bibnamefont{Leylekian}},
  \bibinfo{author}{\bibfnamefont{S.}~\bibnamefont{Ravy}}, \bibnamefont{and}
  \bibinfo{author}{\bibfnamefont{J.~P.} \bibnamefont{Pouget}},
  \bibinfo{journal}{Synth. Metals.} \textbf{\bibinfo{volume}{137}},
  \bibinfo{pages}{1271} (\bibinfo{year}{2003}).

\bibitem[{\citenamefont{D\'ora et~al.}(2002{\natexlab{a}})\citenamefont{D\'ora,
  Virosztek, and Maki}}]{balazs-born}
\bibinfo{author}{\bibfnamefont{B.}~\bibnamefont{D\'ora}},
  \bibinfo{author}{\bibfnamefont{A.}~\bibnamefont{Virosztek}},
  \bibnamefont{and} \bibinfo{author}{\bibfnamefont{K.}~\bibnamefont{Maki}},
  \bibinfo{journal}{Phys. Rev. B} \textbf{\bibinfo{volume}{66}},
  \bibinfo{pages}{115112} (\bibinfo{year}{2002}{\natexlab{a}}).

\bibitem[{\citenamefont{D\'ora et~al.}(2003)\citenamefont{D\'ora, Virosztek,
  and Maki}}]{balazs-unitary}
\bibinfo{author}{\bibfnamefont{B.}~\bibnamefont{D\'ora}},
  \bibinfo{author}{\bibfnamefont{A.}~\bibnamefont{Virosztek}},
  \bibnamefont{and} \bibinfo{author}{\bibfnamefont{K.}~\bibnamefont{Maki}},
  \bibinfo{journal}{Phys. Rev. B} \textbf{\bibinfo{volume}{68}},
  \bibinfo{pages}{075104} (\bibinfo{year}{2003}).

\bibitem[{\citenamefont{V\'anyolos et~al.}()\citenamefont{V\'anyolos, D\'ora,
  Maki, and Virosztek}}]{nca-impurity}
\bibinfo{author}{\bibfnamefont{A.}~\bibnamefont{V\'anyolos}},
  \bibinfo{author}{\bibfnamefont{B.}~\bibnamefont{D\'ora}},
  \bibinfo{author}{\bibfnamefont{K.}~\bibnamefont{Maki}}, \bibnamefont{and}
  \bibinfo{author}{\bibfnamefont{A.}~\bibnamefont{Virosztek}},
  \bibinfo{note}{cond-mat/0606578}.

\bibitem[{\citenamefont{Gr\"uner}(1994)}]{gruner-book}
\bibinfo{author}{\bibfnamefont{G.}~\bibnamefont{Gr\"uner}},
  \emph{\bibinfo{title}{Density waves in solids}}
  (\bibinfo{publisher}{Addison-Wesley}, \bibinfo{address}{Reading},
  \bibinfo{year}{1994}).

\bibitem[{\citenamefont{D\'ora et~al.}(2001)\citenamefont{D\'ora, Virosztek,
  and Maki}}]{balazs-threshold}
\bibinfo{author}{\bibfnamefont{B.}~\bibnamefont{D\'ora}},
  \bibinfo{author}{\bibfnamefont{A.}~\bibnamefont{Virosztek}},
  \bibnamefont{and} \bibinfo{author}{\bibfnamefont{K.}~\bibnamefont{Maki}},
  \bibinfo{journal}{Phys. Rev. B} \textbf{\bibinfo{volume}{64}},
  \bibinfo{pages}{041101(R)} (\bibinfo{year}{2001}).

\bibitem[{\citenamefont{D\'ora et~al.}(2002{\natexlab{b}})\citenamefont{D\'ora,
  Virosztek, and Maki}}]{balazs-magneticthreshold}
\bibinfo{author}{\bibfnamefont{B.}~\bibnamefont{D\'ora}},
  \bibinfo{author}{\bibfnamefont{A.}~\bibnamefont{Virosztek}},
  \bibnamefont{and} \bibinfo{author}{\bibfnamefont{K.}~\bibnamefont{Maki}},
  \bibinfo{journal}{Phys. Rev. B} \textbf{\bibinfo{volume}{65}},
  \bibinfo{pages}{155119} (\bibinfo{year}{2002}{\natexlab{b}}).

\bibitem[{\citenamefont{D\'ora}(2005)}]{balazs-friedel}
\bibinfo{author}{\bibfnamefont{B.}~\bibnamefont{D\'ora}},
  \bibinfo{journal}{Europhys. Lett.} \textbf{\bibinfo{volume}{70}},
  \bibinfo{pages}{362} (\bibinfo{year}{2005}).

\bibitem[{\citenamefont{Clogston}(1962)}]{clogston}
\bibinfo{author}{\bibfnamefont{A.~M.} \bibnamefont{Clogston}},
  \bibinfo{journal}{Phys. Rev.} \textbf{\bibinfo{volume}{125}},
  \bibinfo{pages}{439} (\bibinfo{year}{1962}).

\bibitem[{\citenamefont{Fetter}(1965)}]{fetter}
\bibinfo{author}{\bibfnamefont{A.~L.} \bibnamefont{Fetter}},
  \bibinfo{journal}{Phys. Rev.} \textbf{\bibinfo{volume}{140}},
  \bibinfo{pages}{A1921} (\bibinfo{year}{1965}).

\end{thebibliography}

\end{document}